\documentclass[pre,amsmath,aps,superscriptaddress,twocolumn]{revtex4}
\usepackage{graphicx,xspace}
\usepackage{float}
\usepackage{amsmath}
\usepackage{amssymb}
\usepackage[normalem]{ulem} 
\usepackage{epsfig}
\usepackage{graphicx,psfrag,xspace}

\begin{document}

\title{A nonlinear random walk approach to concentration-dependent contaminant transport in porous media}
\author{Andrea Zoia}
\email{andrea.zoia@cea.fr}
\affiliation{CEA/Saclay, DEN/DM2S/SFME/LSET, B\^at.~454, 91191 Gif-sur-Yvette Cedex, France}
\author{Christelle Latrille}
\affiliation{CEA/Saclay, DEN/DPC/SECR/L3MR, B\^at.~450, 91191 Gif-sur-Yvette Cedex, France}
\author{Alain Cartalade}
\affiliation{CEA/Saclay, DEN/DM2S/SFME/LSET, B\^at.~454, 91191 Gif-sur-Yvette Cedex, France}

\begin{abstract}
We propose a nonlinear random walk model to describe the dynamics of dense contaminant plumes in porous media. A coupling between concentration and velocity fields is found, so that transport displays non-Fickian features. The qualitative behavior of the pollutant spatial profiles and moments is explored with the help of Monte Carlo simulation, within a Continuous Time Random Walk approach. Model outcomes are then compared with experimental measurements of variable-density contaminant transport in homogeneous and saturated vertical columns.
\end{abstract}
\maketitle

\section{Introduction}

Non-Fickian (anomalous) transport is a widespread feature of contaminant migration in porous media~\cite{rev_geo}. Specifically, `non-Fickian' means that the spread of the transported species grows nonlinearly in time, $\langle x^2(t) - \langle x(t) \rangle^2  \rangle \sim t^\beta$, $\beta \ne 1$, the resulting concentration profiles displaying a non-Gaussian behavior~\cite{sahimi, scher_framework, rev_geo}. This is in contrast with the linear spread and Gaussian shapes usually expected for particles migration in perfectly homogeneous media, where the Fickian advection-dispersion equation applies: see, e.g.,~\cite{cortis_homog} and References therein. A broad spectrum of physical reasons have been invoked to explain the observed deviations from Gaussianity. For instance, the homogeneity hypothesis becomes questionable in presence of irregularities at multiple space scales~\cite{fractures, levy}, complex structures of flow streams~\cite{kirchner, zoia} and saturation distribution within the medium~\cite{bromly}, and physico-chemical exchanges of the pollutant particles with the surrounding material~\cite{berkowitz_sorp}. Another important source of non-Fickian behaviors is the collective motion of pollutants due to reciprocal interactions. A well-known example is provided by reactive transport, where two or more chemical species may combine (reversibly or irreversibly) to give birth to new ones. Even in homogeneous media, this may lead to intricate contaminant patterns~\cite{grindrod}, whose complexity could be further increased by the presence of spatial heterogeneities~\cite{havlin}.

Intuitively, the dynamics of concentrated particles will also display nonlinear, collective phenomena. Indeed, the motion of a single pollutant parcel depends on the density of the surrounding fluid, which in turn is affected by the number of such parcels nearby, so that the microscopic trajectories are correlated. Transport of dense pollutant plumes has been long investigated, yet keeps raising many conceptual as well as practical issues~\cite{schincariol, gelhar1, gelhar2, tartakovsky_jfm, tartakovsky_prl, jiao, wood, johannsen, schotting, dangelo, tchelepi}. Studies cover both homogeneous saturated and heterogenous unsaturated materials~\cite{dalziel, oltean, liudane, rogerson}: extensive reviews may be found, e.g., in \cite{rev1, rev2}. Strong density gradients are encountered when either the contaminant itself is highly concentrated at the source, or the plume flows through regions that are rich in salt; in particular, this latter case might become a major concern for radioactive waste disposal near salt domes~\cite{hassanizadeh}.

Similarly as Brownian motion is related to the diffusion equation, concentration-dependent particles paths can be formally shown to lead to a family of nonlinear Fokker-Planck transport equations, on the grounds of a statistical-mechanical approach; see, e.g.,~\cite{chavanisEJP, boon_epl} for a detailed account of recent advances. The displacements of a particle in the medium are thought to be affected by the number of other particles in its initial or final position, or both~\cite{kaniadakis}: this allows better understanding the small-scale dynamics, rather than imposing the macroscopic equations on a phenomenological basis~\cite{chavanisEJP, boon_epl, chavanis, boon_pre, kaniadakis}.

Adopting a somewhat similar perspective, we propose here a simple model for the collective concentration-dependent dynamics of a dense contaminant plume and explore its qualitative behavior by resorting to Monte Carlo simulation. Model predictions are then validated on experimental data. This paper is organized as follows: in Section~\ref{model}, we develop a stochastic equation that describes the motion of a pollutant parcel in a dense fluid. In Section~\ref{discussion}, we discuss the qualitative behavior of the model and the interplay of its components. Then, in Section~\ref{comparison} we proceed to compare model outcomes to experimental results of variable-density contaminant transport in saturated homogeneous porous columns. Finally, the potentialities and the limits of the proposed approach are evidenced in Section~\ref{conclusions}.

\section{A nonlinear transport model}
\label{model}

Let us consider a vertical column filled with sand. For sake of simplicity, we start by assuming that the sand is uniformly packed and well-mixed, so that the porous medium can be considered as homogeneous, and that the column is fully saturated in water. When the ratio between the length and the diameter of the column (the so-called aspect ratio) is much greater than one, the system can be regarded as one-dimensional, to a first approximation. Suppose now that a given amount of contaminant fluid is injected into the column: we can conceptually represent the pollutant plume as a collection of fluid parcels $i=1, ..., N$, each containing a fraction $m_i=M/N$ of the total contaminant mass $M$. When the effects of molecular diffusion are negligible, it is reasonable to assume that $m_i$ will not change in the course of plume evolution~\cite{sph}. If $V$ is the reference volume of the injected pollutant, each parcel carries a volume $v_i=V/N$.

\begin{figure}[t]
\centerline{\epsfxsize=9.0cm\epsfbox{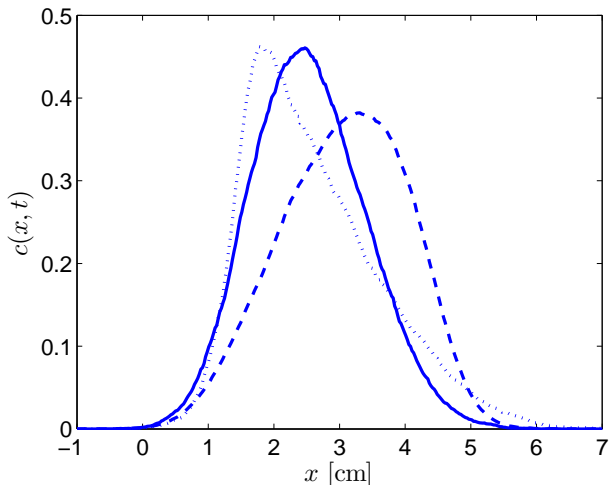}}
\caption{Concentration profiles at time $t=0.45$ h, for step injection from $t=0$ h to $t=0.23$ h. Solid line represents Fickian transport ($\epsilon = 0$); dotted line (injection from the top) and dashed line (injection from the bottom) represent nonlinear concentration-dependent transport.}
   \label{fig1}
\end{figure}

The projections of forces acting on a parcel $i$ in the direction of the flow are: the pressure gradient imposed by the injecting pump, $F_p$, supposedly constant; the viscous resistence which opposes flow, namely $F_v=-\gamma u_i(t)$, where the friction coefficient $\gamma=\mu/k$ is given by the ratio of the fluid dynamic viscosity $\mu$ [Kg/m s] and the medium permeability $k$ [m$^2$], and $u_i(t)$ is the local velocity of a parcel; gravity and buoyancy, which can be written as $F_g=g(\rho_i-\rho_i^f)$, where $g$ is the gravity acceleration, $\rho_i$ is the density of the contaminant parcel and $\rho_i^f$ is the density of the fluid surrounding the parcel $i$. Mechanical dispersion can be taken into account by adding stochastic fluctuations $S_i$ around the parcel velocity $u_i(t)$~\cite{dagan}. It is customary to assume
\begin{equation}
S_i \propto \sqrt{|\langle u_i \rangle |} \eta_i,
\end{equation}
where $\langle u_i \rangle$ is the ensemble average of the particles velocities (provided that the medium is sufficiently homogeneous~\cite{cortis_pores}) and $\eta_i$ is an uncorrelated white noise with zero mean and unit variance. The constant of proportionality determines the strength of the velocity fluctuations and is thus related to the dispersivity $\alpha$ [m] of the porous material. Then, the forces balance reads
\begin{equation}
\rho_i \dot{u}_i  = F_p -\gamma u_i + g(\rho_i-\rho_i^f) + S_i,
\label{langevin}
\end{equation}
where the reference axes system is chosen so that gravity is positive pointing downwards and the explicit dependence on time has been omitted. It appears that the absolute value of the pollutant density does not play a major role, the plume migration being mostly controlled by relative density differences: this is coherent with experimental evidences~\cite{hassanizadeh, simmons, liu, landman1, landman2, landman3}.

\begin{figure}[t]
\centerline{\epsfxsize=9.0cm\epsfbox{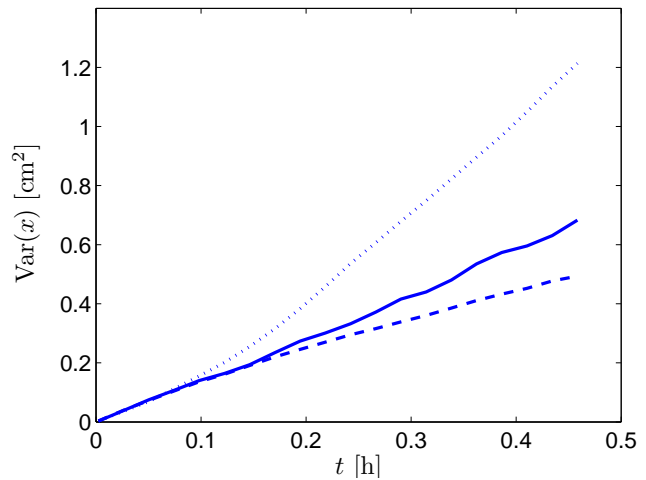}}
\caption{Variance of the particles plume as a function of time, for step injection from $t=0$ h to $t=0.23$ h. Fickian (dots, $\epsilon=0$) and concentration-dependent (dotted line: injection from the top; dashed line: injection from the bottom) transport processes are displayed.}
   \label{fig2}
\end{figure}

Let us now focus on the two terms $\rho_i$ and $\rho_i^f$. The density of a contaminant parcel can be expressed as
\begin{equation}
\rho_i=\frac{\rho_0 v_i + m_i}{v_i},
\end{equation}
where $\rho_0 v_i$ is the mass of reference fluid (e.g., water) contained in $v_i$ and $\rho_0$ its density. By resorting to the definitions of $m_i$ and $v_i$, we obtain
\begin{equation}
\rho_i=\rho_0 \left(1 + \frac{1}{\rho_0} \frac{M}{V} \right) ,
\label{exp1}
\end{equation}
where finally $M/V$ is given by the product of the molar concentration $C^{mol}$ [mol/L] times the molar mass [g/mol] of the injected species. Even modest density differences with respect to the resident fluid (of the order of a few percents) might sensibly affect the contaminant dynamics~\cite{wood, simmons}. Hence, we focus on this case and think of $\rho_i$ as a small perturbation compared to $\rho_0$, i.e., $\epsilon=(M/V)/\rho_0 \ll 1$. As for the local fluid density $\rho_i^f$,
\begin{equation}
\rho_i^f=\frac{\rho_0 dx + m(x_i,t)}{dx},
\end{equation}
where $m(x_i,t)$ is the pollutant mass contained in an elementary volume $dx$ around the position $x_i$ of the parcel $i$. We are assuming that each parcel is aware of the presence of the others only at short range, through the effects of local density variations. Since $m(x_i,t) =n(x_i,t) m_i$, where $n(x_i,t)$ is the number of pollutant parcels in $[x_i,x_i+dx]$ at time $t$, we can finally rewrite
\begin{equation}
\rho_i^f=\rho_0 \left(1 + \epsilon \frac{n(x_i,t)}{N_0} \right),
\label{rho_n}
\end{equation}
where $N_0$ is a dimensionless normalization factor such that $M/V=N_0 m_i/dx$. In practice, $N_0$ expresses the (arbitrary, but sufficiently large) number of contaminant parcels that are initially attributed to each $dx$ to represent the average density $M/V$ at injection. At each time step, the quantity $c(x_i,t)=n(x_i,t)/N_0$ identifies the contaminant concentration at position $x_i$.

\begin{figure}[t]
\centerline{\epsfxsize=9.0cm\epsfbox{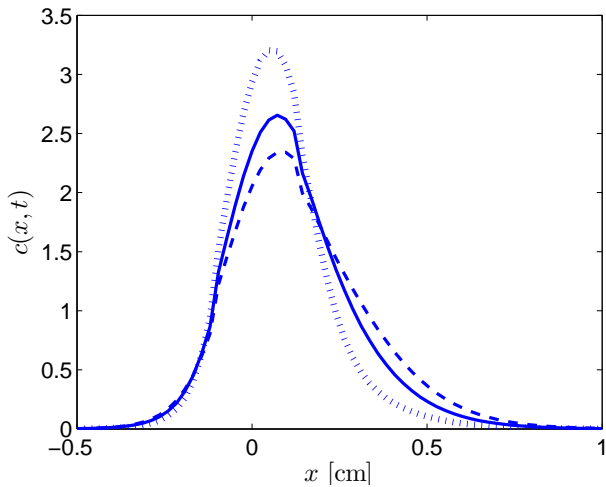}}
\caption{Concentration profiles at time $t=0.48$ h, for step injection from $t=0$ h to $t=0.24$ h. Solid line represents anomalous transport due to spatial heterogeneities, modelled by a waiting times pdf with power-law decay $\psi(\tau) \sim \tau^{-3/2}$ (with $\epsilon = 0$); dotted line (injection from the top) and dashed line (injection from the bottom) represent nonlinear concentration-dependent transport coupled with the effects of the spatial heterogeneities, for the same $\psi(\tau)$.}
   \label{fig3}
\end{figure}

\begin{figure}[t]
\centerline{\epsfxsize=9.0cm\epsfbox{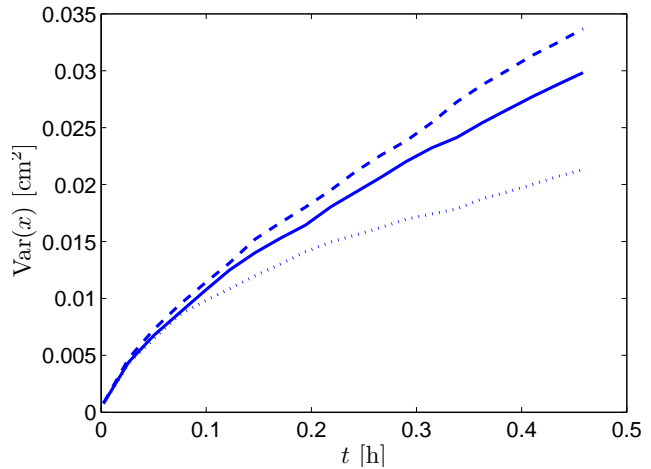}}
\caption{Variance of the particles plume as a function of time, for step injection from $t=0$ h to $t=0.24$ h. Anomalous transport due to spatial heterogeneities, modelled by a waiting times pdf with power-law decay $\psi(\tau) \sim \tau^{-3/2}$ (with $\epsilon=0$) is represented with a solid line. Concentration-dependent transport, coupled with the effects of the spatial heterogeneities, is displayed as dotted line (injection from the top) and dashed line (injection from the bottom), for the same $\psi(\tau)$.}
   \label{fig4}
\end{figure}

The role of viscosity has been condensed in the constant parameter $\gamma$. In reality, viscosity depends on contaminant concentration, but its variations are frequently less relevant than those of density and are thus neglected~\cite{gelhar1, gelhar2}. Within the proposed formulation, including a functional dependence of the kind $\gamma=\gamma_0 (1+\lambda c(x_i,t))$, where $\gamma_0$ is the reference value in the fluid and $\lambda$ is a (small) constant, would be straightforward. In the following, however, we always suppose that $\gamma \simeq \gamma_0$. Moreover, we do not address the possible dependence of density and viscosity on other physical variables, such as temperature.

Finally, assuming that inertial effects can be neglected (which is the case, provided that viscous forces are dominant), and making use of expressions~\ref{exp1} and~\ref{rho_n}, we can rewrite Eq.~\ref{langevin} in Langevin form
\begin{equation}
\dot{x}_i = u(c) + \gamma^{-1} S_i.
\label{law}
\end{equation}
Equation~\ref{law} describes the random walk of a fluid parcel which is advected at a concentration-dependent speed $u(c)=u_p+u_g+u_c$, with $u_p=\gamma^{-1} F_p$, $u_g=\gamma^{-1} g \rho_0 \epsilon$ and $u_c=-\gamma^{-1} g \rho_0 \epsilon c(x_i,t)$, and dispersed by fluctuations whose amplitude is $std (\gamma^{-1} S_i dt) = \left[ 2 \alpha | \langle u_i(t) \rangle | dt \right]^{1/2}$. Note that dispersion $D(c)=\alpha | \langle u_i(t) \rangle |$ is also a function of concentration, through the dependence on the ensemble-averaged velocity.

By relying upon the results resumed in, e.g.,~\cite{chavanis}, it is possible to show that the smoothed contaminant concentration field $c(x,t)=\langle \sum_i \delta(x-x_i(t)) \rangle$ corresponding to particles undergoing the random walk in~\ref{law} obeys a nonlinear Fokker-Planck equation
\begin{equation}
\frac{\partial }{\partial t} c(x,t)=-\frac{\partial }{\partial x} \left[ u(c(x,t)) -\frac{\partial }{\partial x} D(c(x,t)) \right] c(x,t).
\label{nonlinfokker}
\end{equation}
Equations of this form are well-known and commonly arise in the context of transport processes with concentration-dependent dispersion and/or velocity: see, e.g.,~\cite{transp_phen}. While we will not make explicit use of its properties in the following, Eq.~\ref{nonlinfokker} provides the necessary link between the microscopic stochastic particles dynamics in Eq.~\ref{law} and the deterministic evolution of the associated ensemble-averaged concentration field. Note that the effects of mutual interactions in Eq.~\ref{nonlinfokker} become negligible for $\epsilon \to 0$, i.e., when the molar concentration of the injected solution is weak. In this case, the particles trajectories are independent, $u_i(t) \to u_p$, and Eq.~\ref{nonlinfokker} degenerates to a standard advection-dispersion equation, so that Fickian transport is recovered, with $D=\alpha u_p$. For a given value of $\epsilon >0$, the nonlinear coupling plays a minor role at short time scales also for $n(x_i,t) \to 0$, i.e., when the number of contaminant particles in the considered $dx$ is small. This is the case when dispersion dominates, so that fluid parcels are rapidly dragged far apart and can hardly interact. Eventually, at longer time scales, dispersion will usually overcome the effects due to concentration.

\section{Discussion}
\label{discussion}

Equation~\ref{law} defines a discrete-time random walk where the particles positions are updated at each time step $dt$. In view of the possibility of describing a broad class of porous materials, such as heterogeneous and/or unsaturated media, it is expedient to resort to the more general Continuous Time Random Walk (CTRW) formalism~\cite{silbey, rev_geo}, where particles trajectories alternate random jumps (drawn from a pdf $p(s)$) and random waiting times (drawn from a pdf $\psi(\tau)$) at each visited spatial site. The pdf $\psi(\tau)$ identifies the velocity spectrum in the traversed material: flows in homogeneous porous media such as those considered here (where it is reasonable to assume that the sojourn times at each site must be on average the same~\cite{rev_geo}) correspond to choosing a Poisson pdf $\psi(\tau)$, so that a single time-scale, e.g., the average $\langle \tau \rangle$ of the distribution, dominates~\cite{rev_geo}. As for the displacements, the spatial scales of advection and dispersion in the CTRW are determined by the cumulants $\kappa$ of the jump lengths distribution $p(s)$ and are not separated {\em a priori}~\cite{rev_geo}. A common choice is to adopt a Gaussian pdf $p(s)$, so that the first two cumulants are sufficient to characterize transport: in particular, $\kappa_1$ is associated to advection and $\kappa_2$ to dispersion.

\begin{figure}[t]
\centerline{\epsfxsize=9.0cm\epsfbox{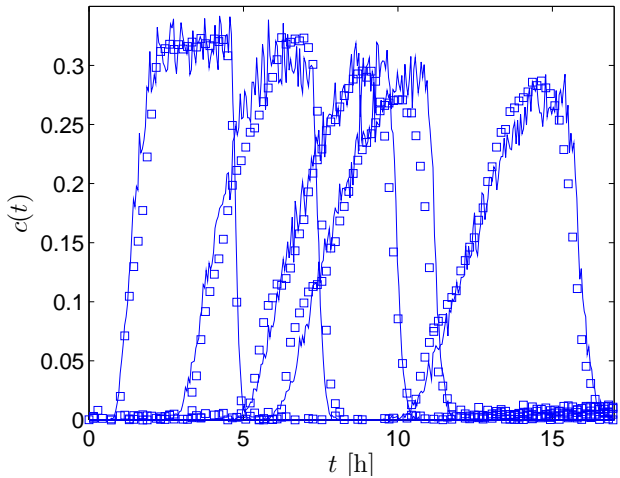}}
\caption{Downwards injection at a reference molarity $C^{mol}=0.2$ mol/L. Contaminant concentration curves $c_\ell(t)$ measured at sections $\ell =7.7, 23.1, 38.5, 46.2,$ and $77$ cm (from left to right), as a function of time. Squares correspond to experimental data, solid lines to Monte Carlo simulation.}
   \label{fig5}
\end{figure}

\begin{figure}[t]
\centerline{\epsfxsize=9.0cm\epsfbox{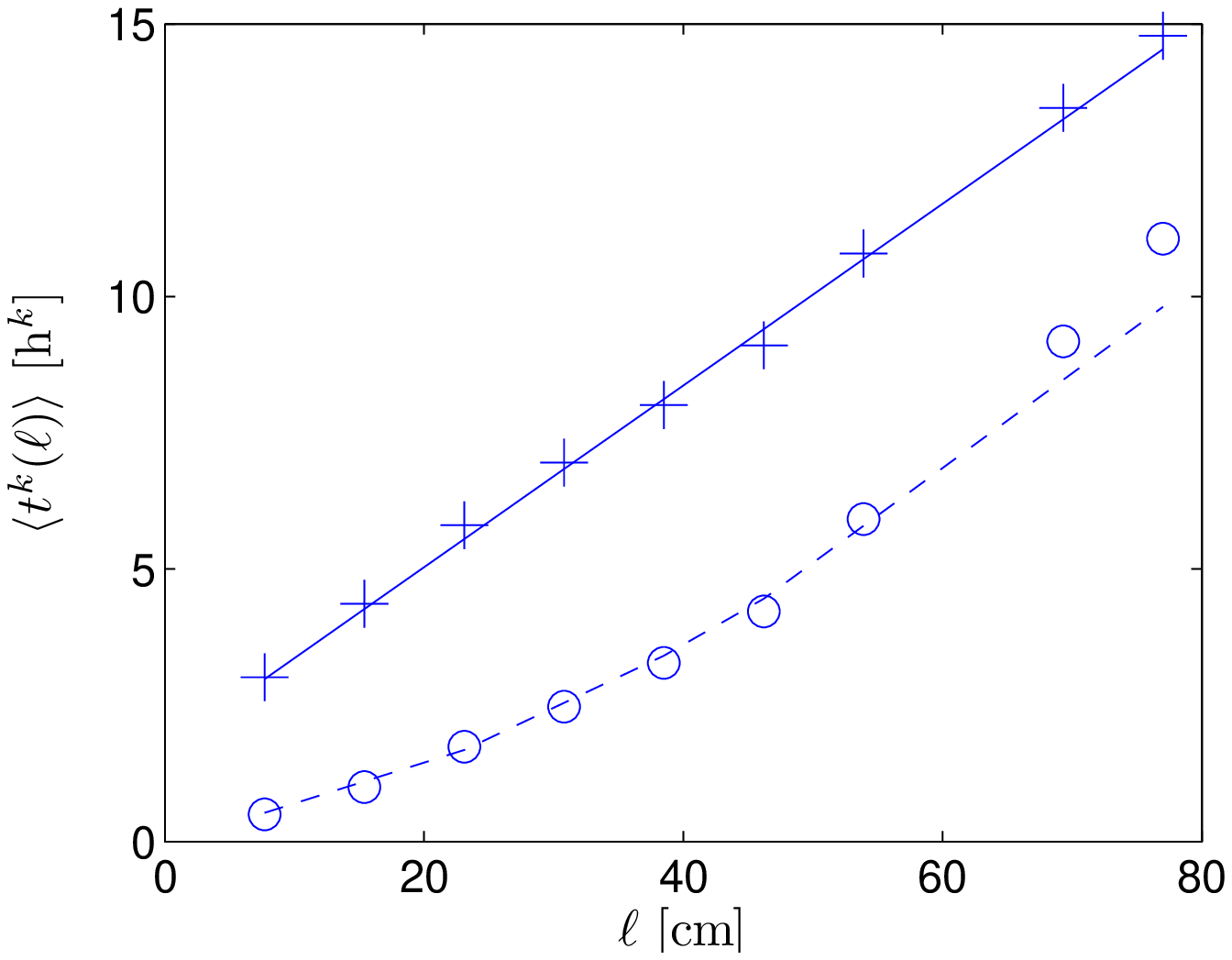}}
\caption{Downwards injection at a reference molarity $C^{mol}=0.2$ mol/L. Moments $\langle t^k(\ell)\rangle$ of passage times $t(\ell)$, as a function of various column heights $\ell$. Crosses represent the mean of the passage times ($k=1$), circles the second moment ($k=2$); the latter has been divided by a factor of $20$ in order to have comparable scales. Solid ($k=1$) and dashed lines ($k=2$) are the results of Monte Carlo simulation.}
   \label{fig6}
\end{figure}

We can then rephrase the stochastic process defined in Eq.~\ref{law} by resorting to a CTRW where waiting times obey a Poisson pdf with mean $\langle \tau \rangle$, i.e.,
\begin{equation}
\psi(\tau)= \frac{1}{\langle \tau \rangle} e^{-\tau/\langle \tau \rangle}
\label{mc1}
\end{equation}
and jumps obey a Gaussian pdf with concentration-dependent cumulants $\kappa_1=u(c) \langle \tau \rangle$ and $\kappa_2=2D(c)\langle \tau \rangle$, i.e.,
\begin{equation}
p(s) = p(s|c)= \frac{1}{\sqrt{2 \pi \kappa_2}} e^{-(s-\kappa_1)^2 / 2 \kappa_2}.
\label{mc2}
\end{equation}
The CTRW formalism has been here introduced on phenomenological basis, as a generalization of Eq.~\ref{law}; hints for a rigorous derivation of concentration-dependent transition rates within a nonlinear master equation formulation are provided, e.g., in~\cite{kaniadakis}. The process defined by~\ref{mc1} and~\ref{mc2} can be easily simulated by Monte Carlo method. Note however that the jump lengths distribution $p(s|c)$ explicitly depends on concentration, so that particles trajectories are not independent and mutual interactions require knowing the concentration field (i.e., the locations of the entire ensemble) before updating walkers positions. Starting from a known initial condition $c(x,0)$, particles are displaced at each time step by drawing waiting times and jump lengths from~\ref{mc1} and~\ref{mc2}, respectively, and the new concentration field is recursively determined for the following time step. Walkers whose random waiting time is longer than the current time step stay in the same spatial site. For continuous contaminant spills, new particles are added at the column entrance for the duration of the injection. In order to attain convergence in simulations, we could either $i)$ choose very small time steps for updating displacements and concentration field, or $ii)$ at each (larger) time step iteratevely compute the values of displacements and concentration until their relative error is below a given threshold. After testing both methods, we found convenient to resort to the former: the optimal value of the time step was determined by trial-and-error.

Monte Carlo simulation offers an expedient means of exploring the qualitative features of the nonlinear CTRW transport model described above. In particular, we proceed now to analyze spatial contaminant concentration profiles (at fixed time) for small values of the parameter $\epsilon$; this allow getting insights on the relevance of the coupling between velocity and concentration. Figure~\ref{fig1} compares a Fickian contaminant profile, corresponding to $\epsilon=0$, with typical spatial profiles for concentration-dependent transport ($0 < \epsilon \ll 1$). Downwards injection gives rise to positively skewed profiles, while the opposite is true for upwards injection. In all cases, we considered a step injection of finite duration. Nonlinear transport clearly displays asymmetric profiles, whereas Fickian transport corresponds to Gaussian (symmetric) profiles.

\begin{figure}[t]
\centerline{\epsfxsize=9.0cm\epsfbox{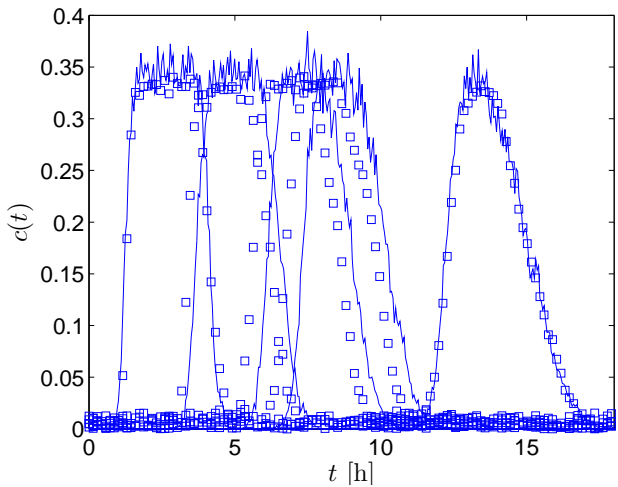}}
\caption{Upwards injection at a reference molarity $C^{mol}=0.1$ mol/L. Contaminant concentration curves $c_\ell(t)$ measured at sections $\ell =7.7, 23.1, 38.5, 46.2,$ and $77$ cm (from left to right), as a function of time. Squares correspond to experimental data, solid lines to Monte Carlo simulation.}
   \label{fig7}
\end{figure}

\begin{figure}[t]
\centerline{\epsfxsize=9.0cm\epsfbox{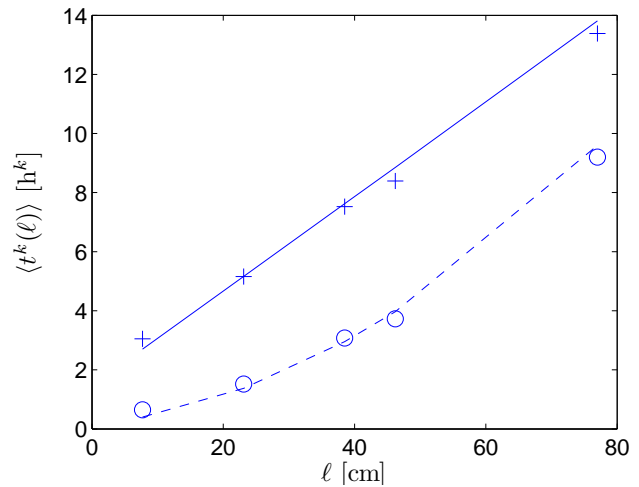}}
\caption{Upwards injection at a reference molarity $C^{mol}=0.1$ mol/L. Moments $\langle t^k(\ell)\rangle$ of passage times $t(\ell)$, as a function of various column heights $\ell$. Crosses represent the mean of the passage times ($k=1$), circles the second moment ($k=2$); the latter has been divided by a factor of $20$ in order to have comparable scales. Solid ($k=1$) and dashed lines ($k=2$) are the results of Monte Carlo simulation.}
   \label{fig8}
\end{figure}

The time duration of the contaminant injection is a key factor in determining the spatial shape of the plume. Indeed, the coupling between concentration and velocity is in competition with dispersion, which in turn is induced by the average velocity $\langle u_i(t)\rangle$. The stronger the velocity $u_p$, the lesser is the relevance of the nonlinear term in $u(c)$. The injected plume might have such a limited extension that dispersion rapidly dominates concentration-dependent effects: in other words, because of their velocity, fluid parcels become quickly dispersed, and their interactions through the density field are weak. This prediction is coherent with our experimental measures: increasing the imposed flux (at fixed $C^{mol}$), the contaminant profiles approach standard Fickian shapes. At the opposite, the longer the extension of the injected plume and the more persistent are the effects of the reciprocal interactions, before eventually dispersion takes over. This phenomenon has already been experimentally detected for the case of viscosity-dependent transport of `slices' of finite duration~\cite{dewit}. Therefore, for a given value of $\epsilon$, the relevance of the nonlinear coupling is stronger for small velocity fields $u_p$ and long injection times.

On the basis of the observations above, one might expect that the effects of the nonlinear coupling would come into play mainly through velocity variations. Actually, it turns out that the average particle velocity $\langle u_i(t) \rangle$ is only slightly affected by density, provided that $\epsilon$ is not too large. On the other hand, fluctuations around the mean velocity (induced by the nonlinear terms) do not simply average out, and contribute instead to an apparent plume dispersion (in addition to $\alpha | \langle u_i(t) \rangle | \simeq \alpha u_p$). This is a relevant and subtle outcome, which is ultimately responsible for the skewed shape of the pollutant profiles, the concentration-dependent contribution to dispersion being proportional to density differences (and thus non-symmetric).

In Fig.~\ref{fig2} we display the behavior of the contaminant variance $\langle x^2(t) - \langle x(t) \rangle^2  \rangle$ as a function of time, as computed by Monte Carlo simulation. For the case of Fickian transport ($\epsilon$=0), the variance is a straight line, as expected. For concentration-dependent transport, the variance turns out to be a nonlinear function of time and appreciably deviates from the Fickian behavior. This is indeed the hallmark of anomalous diffusion. On the contrary, the average of the contaminant plume (not shown here) is found to be linear in time, for small values of $\epsilon$.

\begin{figure}[t]
\centerline{\epsfxsize=9.0cm\epsfbox{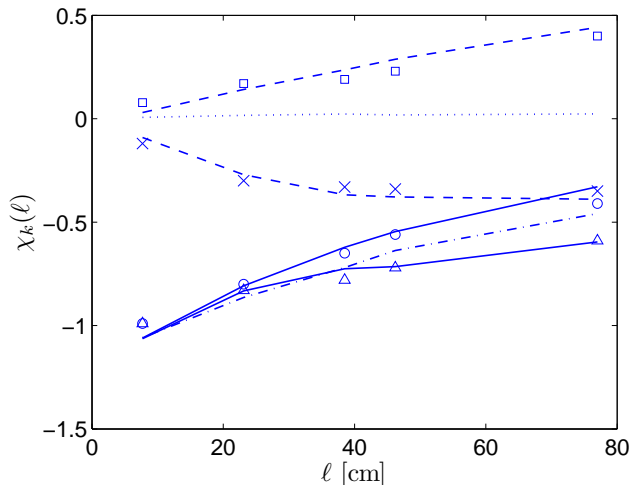}}
\caption{Skewness $\chi_3(\ell)$ and kurtosis $\chi_4(\ell)$ of the arrival times considered in Figs.~\ref{fig5} and~\ref{fig7}, as a function of column height $\ell$. Estimates of $\chi_3(\ell)$ for experimental data: squares (upwards injection) and crosses (downwards); Monte Carlo estimates: dashed lines. Estimates of $\chi_4(\ell)$ for experimental data: circles (upwards injection) and triangles (downwards); Monte Carlo estimates: solid lines. Estimates for Gaussian transport ($\epsilon=0$): dotted ($\chi_3(\ell)$) and dotted-dashed ($\chi_4(\ell)$) lines.}
   \label{fig9}
\end{figure}

While in the present discussion we have made the assumption of considering flows in homogeneous porous media, which amounts to drawing waiting times from a Poisson pdf, the CTRW framework straightforwardly allows taking into account spatial heterogeneities, due, e.g., to different grain sizes or variable saturation. The broad velocities spectra that are commonly found in heterogeneous and/or unsaturated materials are mirrored in a broad distribution of time scales for the jumping rates between sites: it is customary to incorporate such physical processes in $\psi(\tau)$ by considering power-law waiting times between consecutive displacements, possibly with an exponential cut-off~\cite{rev_geo, cortis_pores, cortis_homog}. Particles trajectories would then be affected on one hand by concentration-dependent displacements and on the other hand by anomalously long sojourns: the two processes may superpose, depending on the respective time scales. The qualitative behavior of the competition between density effects and heterogeneities is displayed in Figs.~\ref{fig3} and~\ref{fig4}, where we show spatial contaminant profiles and particles variance, respectively, for a waiting times pdf $\psi(\tau) \sim \tau^{-3/2}$. In particular, we remark that the asymmetry that was evident for homogeneous transport (Fig.~\ref{fig1}) is now hidden by the long tails of the pollutant profiles.

\section{Comparison with experimental results}
\label{comparison}

In this Section, we test the proposed random walk model on some experimental measurements of dense contaminant transport obtained at the Physical-Chemistry Department (DPC), CEA/Saclay. The experimental device, named BEETI, consists of a dichromatic X-ray source ($20-40$ keV, $50-75$ keV), applied to a vertical column of height $H=80$ cm and diameter $D=5$ cm (the aspect ratio is therefore $H/D=16 \gg 1$). The X-ray transmitted countings allow quantitatively assessing the contaminant concentration inside the column (as a function of time), at various sections $\ell$: we denote this quantity by $c_\ell(t)$. The different positions are explored by means of a remotely controlled rack rail that displaces the X-ray emitter and the coupled NaI detector. At the exit of the column, $c_{\ell=H}(t)$ coincides with the breakthrough curve, which is the most frequently measured variable in contaminant migration experiments~\cite{rev_geo}. In the specific context of dense contaminant transport, only a few works have investigated the behavior of breakthrough curves corresponding to finite-duration injections, whereas attention is usually focused on the mixing properties at the interface between two layers of semi-infinite extension (see, e.g.,~\cite{wood, dewit} and References therein).

The BEETI experimental setup allows for downwards as well as upwards fluid injection, and several kinds of flow regimes and porous materials can be tested, at various saturation and/or heterogeneity conditions. To set the ideas, in the following we refer to fully saturated columns filled with homogeneously mixed Fontainebleau sand (bulk density $1.77 \pm 0.01$ g/cm$^3$), with average grain diameter $200$ $\mu$m. The average porosity is $\theta =0.333 \pm 0.005$ and the dispersivity is $\alpha=0.1$ cm. The reference saturating fluid is water containing dissolved KCl (molar mass equal to $74.5$ g/mol) at a molar concentration of $10^{-3}$ mol/L, so that $\rho_0=998.3$ Kg/m$^3$ at $T=20$ C$^{o}$. The injected contaminant is KI (molar mass equal to $166$ $g/mol$), at different molar concentrations. All measurements are performed at constant room temperature $T=20$ C$^{o}$. We estimated $\gamma^{-1} g \rho_0 \simeq 5$ cm/h. Contaminant flow is imposed at one end of the column and collected at the other end, where an electric conductivity meter provides a supplementary (independent) measurement of the breakthtrough curve. The pump imposes a steady state Darcy flow of $q=u_p \theta=2$ cm/h, which is verified by weighing the outgoing solution. The experimental conditions are such that clogging or formation of colloidal particles, which could alter the interpretation of the obtained results, can be excluded. Chemical reactions or sorption/desorption phenomena can be ruled out as well.

A representative example is shown in Figs.~\ref{fig5} and~\ref{fig6} for downwards injection of KI at $q=2$ cm/h, with $C^{mol}=0.2$ mol/L, so that $\epsilon=0.033$. The time duration of injection is $3$ h. Figure~\ref{fig5} compares the experimental concentration profiles (squares) with the Monte Carlo simulation results (solid lines). From the point of view of Monte Carlo simulation, the quantity $c_\ell(t)$ is estimated as the number of particles that are contained in a volume $dx$ around the position $\ell$, at a given time $t$. In other words, $c_\ell(t)$ represents the distribution of the passage times at fixed positions. In principle, knowledge of the physical constants completely determines the free parameters of the simulation; in practice, however, a trial-and-error fine fitting around $\epsilon$ and $\alpha$ is required in order to account for uncertainties. Despite the many assumptions and simplifications introduced in the random walk model, a good agreement is found between simulation and data. This agreement, moreover, is preserved all along the measurement points $\ell$, thus meaning that the proposed model allows capturing the full spatial dynamics of the plume. It is evident that the asymmetric spatial shape that had been predicted on the basis of random walk simulations (Fig.~\ref{fig1}) is now mirrored in the shape of $c_\ell(t)$. Due to the interplay of concentration and velocity, a part of the contaminant plume is descending faster than the bulk.

The agreement between model and experimental data is further substantiated by Fig.~\ref{fig6}, where we compare the first two moments $\langle t^k(\ell)\rangle$, $k=1,2$, of the passage times $t(\ell)$ along the column. We remark that the slope of $\langle t^1(\ell)\rangle$ is very close to the value $u_p^{-1}$, which is consistent with the average particles velocity being almost unaffected by the concentration field. These findings are coherent with experimental observations and models of density-dependent transport proposed in literature~\cite{gelhar1, gelhar2, schotting, hassanizadeh, liu, landman1, landman2, landman3}.

Comparable results have been obtained also for upwards injection. A representative example is shown in Figs.~\ref{fig7} and~\ref{fig8} for $q=2$ cm/h and $C^{mol}=0.1$ mol/L, so that $\epsilon=0.017$. The time duration of injection is $3$ h. The asymmetric tail of the contaminant concentration profiles is now on the right, meaning that part of the bulk is delayed because of density effects (cf.~Fig.~\ref{fig7}). A slightly less satisfactory agreement is found for the profiles at intermediate heights, which could be attributed to neglecting inertial contributions in Eq.~\ref{langevin}. Nonetheless, the breakthrough curve and the moments (Fig.~\ref{fig8}) are well captured by the random walk model.

Finally, in order to emphasize the departure of the concentration profiles shown in~Fig.~\ref{fig5} and~\ref{fig7} from Gaussian behavior, in Fig.~\ref{fig9} we provide the skewness $\chi_3(\ell)$ and kurtosis $\chi_4(\ell)$ of the arrival times~\footnote{$\chi_3(\ell) = \langle (t(\ell)-\langle t(\ell) \rangle)^3 \rangle / \langle (t(\ell)-\langle t(\ell) \rangle)^2 \rangle^{3/2}$; $\chi_4(\ell) = \langle (t(\ell)-\langle t(\ell) \rangle)^4 \rangle/ \langle (t(\ell)-\langle t(\ell) \rangle)^2 \rangle^2 -3  $.}, as a function of column height $\ell$. Experimental data estimates lie close to those of Monte Carlo simulations. For comparison, the case of Gaussian transport (i.e., $\epsilon=0$) is plotted in the same figure: the difference with respect to dense contaminant transport is clearly noticeable. Remark in particular that for downwards injection $\chi_3(\ell)<0$ and decreases with $\ell$, whereas for upwards injection $\chi_3(\ell)>0$ and increases with $\ell$. For $\chi_4(\ell)$, similar deviations from Gaussian behavior are observed, though partially hidden by limited statistics.

In principle, one might wonder whether a standard linear CTRW with algebraic $\psi(t)$, which also gives rise to asymmetric breakthrough curves with long tails, could be applied to fit the experimental data. However, this hypothesis is in contrast with two basic facts: first, adopting a power-law waiting time pdf is somehow unjustified, since the medium is homogeneous; second, a standard linear CTRW approach could not explain why the asymmetry of the breakthrough curves is affected by the flow direction. So far, our experimental activities have exclusively concerned the transport of dense contaminant plumes in homogeneous saturated columns. However, further tests are in order, to explore the case of heterogeneous and/or unsaturated porous media. The BEETI device, thanks to the dual-energy source, can determine at the same time contaminant concentration and water content at each section: it would be thus interesting to compare model predictions (Figs.~\ref{fig3} and~\ref{fig4}) with experimental data.

\section{Conclusions}
\label{conclusions}

We have proposed a nonlinear random walk approach to the modelling of variable-density contaminant flows in porous media, within a CTRW framework. The qualitative behavior of this model has been explored by means of Monte Carlo simulation: particles trajectories are correlated via the density field, so that transport is non-Fickian and the plume variance grows nonlinearly in time. When the molar concentration of the injected pollutants is similar to that of the resident fluid, the usual Fickian behavior is recovered. Within CTRW, it is possible to describe transport through both homogeneous and heterogeneous materials: in this latter case, we have shown that the effects of concentration-dependent dynamics are in competition with (and might partially hidden by) those of spatial heterogeneities.

The proposed random walk model is admittedly simple, since the full spectrum of interactions that actually take place between the velocity and density fields has been condensed in a single nonlinear coupling at the scale of particles trajectories. Detailed studies show that the physics behind variable-density transport is essentially $3d$, or at least $2d$, because of the complex interfacial dynamics between two fluids of different densities and/or viscosities~\cite{gelhar1, tartakovsky_prl, jiao, wood, johannsen, schotting, dangelo, tchelepi}. Neglecting these phenomena leads to descriptions that must be necessarily intended in a mean-field sense: only the coarse-scale behavior of the real system can be captured, and the fine-scale details are averaged out~\cite{tardy, schotting, hassanizadeh, liu, landman1, landman2, landman3}. Moreover, we have made the hypothesis that molecular diffusion is negligible with respect to mechanical dispersion, and that viscosity can be considered as constant, to a first approximation.

Yet, our random walk model compares well to a set of dense contaminant transport measurements realized by means of the BEETI device. The experimental conditions ensure that most of the introduced simplifications actually apply: the aspect ratio of the column is large, so that migration is almost $1d$; viscosity variations are weaker than density variations; molecular diffusion is smaller than dispersion. It seems reasonable to think that the limits of validity of the proposed model will clearly emerge when these hypotheses are not verified: experimental activities are ongoing and will be presented in a forthcoming work. In particular, we expect our model to provide a satisfactory agreement with measured data when it is possible to consider density variations as small perturbations with respect to the resident fluid (i.e., $\epsilon \ll 1$). For larger density differences, other, more complex couplings should perhaps be introduced, possibly involving higher-order nonlinearities and long-range correlations. The findings in~\cite{hassanizadeh}, for instance, suggest that in presence of relevant density gradients even the validity of Fick and Darcy laws at microscopic scale should be carefully reconsidered. In this respect, Monte Carlo simulation might be complemented, e.g., by the promising computational tool of Smoothed Particle Hydrodynamics, which has been recently applied with success to the numerical study of variable-density flows with stochastic dispersion~\cite{tartakovsky_prl}.

The proposed random walk approach has been motivated by a specific problem in contaminant migration; many other physical processes where the CTRW formalism applies may exhibit particles paths correlated via the density field, so that relevant advances would be achieved by formally generalizing the CTRW theory for the case of concentration-dependent distributions.\newline

\acknowledgments
A.Z.~thanks Ph.~Montarnal, Ph.~Roblin (CEA/Saclay) and A.~Cortis (Lawrence Berkeley National Laboratory) for useful discussions and comments.


\begin{thebibliography}{0}

\bibitem{rev_geo} B. Berkowitz, A. Cortis, M. Dentz, and H. Scher, Rev. Geophys. {\bf 44}, RG2003 (2006). 

\bibitem{sahimi} M. Sahimi, {\em Flow and Transport in Porous Media and Fractured Rock} (VCH, Weinheim 1995).

\bibitem{scher_framework} H. Scher, G. Margolin, and B. Berkowitz, Chem. Phys. {\bf 284}, 349 (2002). 

\bibitem{cortis_homog} A. Cortis and B. Berkowitz, Soil Sci. Soc. Am. J. {\bf 68}, 1539 (2004). 

\bibitem{fractures} B. Berkowitz and H. Scher, Phys. Rev. Lett. {\bf 79}, 4038 (1997). 

\bibitem{levy} M. Levy and B. Berkowitz, J. Contam. Hydr. {\bf 64}, 203 (2003). 

\bibitem{kirchner} J. W. Kirchner, X. Feng, and C. Neal, Nature {\bf 403}, 524 (2000). 

\bibitem{zoia} A.~Zoia, Y.~Kantor, and M.~Kardar, EuroPhys. Lett. {\bf 80}, 40006 (2007). 

\bibitem{bromly} M. Bromly and C. Hinz, Water Resour. Res. {\bf 40}, W07402 (2004). 

\bibitem{berkowitz_sorp} B. Berkowitz, S. Emmanuel, and H. Scher, Water Resour. Res. {\bf 44}, W03402 (2008). 

\bibitem{grindrod} P. Grindrod, {\em Patterns and waves: The theory and applications of reaction-diffusion equations} (Clarendon Press, 1991).

\bibitem{havlin} D. ben-Avraham and S. Havlin, {\em Diffusion and reactions in fractals and disordered systems} (Cambridge University Press, Cambridge, UK, 2005).

\bibitem{schincariol} R. A. Schincariol and F. W. Schwartz, Water Resour. Res. {\bf 26}, 2317 (1990). 

\bibitem{gelhar1} C. Welty and L. W. Gelhar, Water. Resour. Res. {\bf 27}, 2061 (1991). 

\bibitem{gelhar2} C. Welty and L. W. Gelhar, Water. Resour. Res. {\bf 28}, 815 (1992). 

\bibitem{tchelepi} H. A. Tchelepi, F. M. Orr, Jr., N. Rakotomalala, D. Salin, and F. L. Woum\'eni, Phys. Fluids A {\bf 5} (7) (1993).

\bibitem{schotting} R. J. Schotting, H. Moser, and S. M. Hassanizadeh, Adv. Water Resour. {\bf 22}, 665 (1999). 

\bibitem{wood} M. Wood, C. T. Simmons, and J. L. Hutson, Water Resour. Res. {\bf 40}, W03505 (2004). 

\bibitem{jiao} C.-Y. Jiao and H. Hotzl, Transp. Porous Media {\bf 54}, 125 (2004). 

\bibitem{tartakovsky_jfm} M. Dentz, D. M. Tartakovsy, E. Abarca, A. Guadagnini, X. Sanchez-Vila, and J. Carrera, J. Fluid. Mech. {\bf 561}, 209 (2006). 

\bibitem{johannsen} K. Johannsen, S. Oswald, R. Held, and W. Kinzelbach, Adv. Water Resour. {\bf 22}, 1690 (2006). 

\bibitem{tartakovsky_prl} A. M. Tartakovsky, D. M. Tartakovsky, and P. Meakin, Phys. Rev. Lett. {\bf 101}, 044502 (2008).

\bibitem{dangelo} M. V. D'Angelo, H. Auradou, C. Allain, M. Rosen, and J.-P. Hulin, Phys. Fluids {\bf 20}, 034107 (2008). 

\bibitem{dalziel} S. B. Dalziel, M. D. Patterson, C. P. Caulfield, and I. A. Coomaraswamy, Phys. Fluids {\bf 20}, 065106 (2008). 

\bibitem{oltean} C. Oltean and M. A. Bu\'es, Transp. Porous Media {\bf 48}, 61 (2002). 

\bibitem{liudane} H. H. Liu and J. H. Dane, J. Hydrology {\bf 194}, 126 (1997). 

\bibitem{rogerson} A. Rogerson and E. Meiburg, Phys. Fluids A {\bf 5} (11), (1993). 

\bibitem{rev1} C. T. Simmons, T. R. Fenstemaker, and J. M. Sharp Jr., J. Contam. Hydrology {\bf 52}, 245 (2001). 

\bibitem{rev2} H.-J. G. Diersch and O. Kolditz, Adv. Water Resour. {\bf 25}, 899 (2002). 

\bibitem{hassanizadeh} S. M. Hassanizadeh and A. Leijnse, Adv. Water Resour. {\bf 18}, 203 (1995). 

\bibitem{chavanisEJP} P. H. Chavanis, Eur. Phys. J. B {\bf 62}, 179 (2008). 

\bibitem{boon_epl} J. P. Boon and J. F. Lutsko, EuroPhys. Lett. {\bf 80}, 60006 (2007). 

\bibitem{kaniadakis} G. Kaniadakis, Physica A {\bf 296}, 405 (2001). 

\bibitem{chavanis} P. H. Chavanis, Phys. Rev. E {\bf 68}, 036108 (2003). 

\bibitem{boon_pre} J. F. Lutsko and J. P. Boon, Phys. Rev. E {\bf 77}, 051103 (2008). 

\bibitem{sph} A. M. Tartakovsky and P. Meakin, J. Comp. Physics {\bf 207}, 610 (2005). 

\bibitem{dagan} G. Dagan and S. P. Neuman (Eds.), {\em Subsurface flow and transport: A stochastic approach} (Cambridge University Press, Cambridge, UK, 2005).

\bibitem{cortis_pores} A. Cortis, Y. Chen, H. Scher, and B. Berkowitz, Phys. Rev. E {\bf 70}, 041108 (2004). 

\bibitem{simmons} C. T. Simmons, M. L. Pierini and J. L. Hutson, Transp. Porous Media {\bf 47}, 215 (2002). 

\bibitem{liu} H. H. Liu and J. H. Dane, Transp. Porous Media {\bf 23}, 219 (1996). 

\bibitem{landman1} A. J. Landman, K. Johannsen, and R. Schotting, Adv. Water Resour. {\bf 30} 2467 (2007). 

\bibitem{landman2} A. J. Landman, R. Schotting, A. Egorov, and D. Demidov, Adv. Water Resour. {\bf 30}, 2481 (2007). 

\bibitem{landman3} A. G. Egorov, D. E. Demidov, and R. Schotting, Adv. Water Resour. {\bf 28}, 55 (2005). 

\bibitem{transp_phen} R. B. Bird, W. E. Stewart, and E. N. Lightfoot, {\em Transport phenomena} (J. Wiley $\&$ Sons., New York, USA, 2005).

\bibitem{silbey} J. Klafter and R. Silbey, Phys. Rev. Lett. {\bf 44}, 55 (1980). 

\bibitem{dewit} A. De Wit, Y. Bertho, and M. Martin, Phys. Fluids {\bf 17}, 054114 (2005). 

\bibitem{tardy} P. M. J. Tardy and J. R. A. Pearson, Transp. Porous Media {\bf 62}, 205 (2006). 

\end{thebibliography}
\end{document}